# Dynamics of center-periphery patterns in knowledge networks – the case of China's biotech science and technology system

Stefan HENNEMANN[1], Tao WANG[2], Ingo LIEFNER[1]

**Abstract**

Science and technology systems – and their epistemic communities – are usually hierarchical and composed of a number of strong, large, leading organizations, along with a number of smaller and less influential ones. Moreover, these hierarchical patterns have a spatial structure: the leading organizations are concentrated in a few places, creating a science and technology center, whereas the majority of locations are peripheral. In the example of biotech research in China, we found dynamic changes in center-periphery patterns. These results are based on a network analysis of evolving co-authorship networks from 2001 to 2009 that were built combining national and international databases.

Therefore, our results are not only relevant for evaluating the spatial structure and dynamics in the Chinese biotech system and its integration into the global knowledge network, but also revive a discussion on persistence and processes of change in the systems theory for science-based industries.

[1] Justus-Liebig-University Giessen, Economic Geography, Senckenbergstrasse 1, D-35390, GERMANY
[2] Nanjing Normal University, School of Geographical Sciences, 1 Wenyuan Road, Nanjing, CHINA

# 1 Introduction

The biotech industry, and the research activity in this science-driven sector in particular, is seen as being highly concentrated in several global spots and largely organized in epistemic communities that act on a global scale, rather than creating and utilizing 'local buzz' (Cooke, 2009; Moodysson, 2008). Consequently, knowledge dissemination processes are assumed to be organized in a hierarchical fashion, with the most highly ranked organizations circulating knowledge among themselves and, at later stages, passing over knowledge to subordinate levels (cf. Hennemann, 2010). The basic logic behind this stratified diffusion is that of the different ability to produce, absorb and disseminate knowledge among the various players. Leading organizations such as global companies or global universities are spearheading the knowledge production, while subordinate tiers of the knowledge production system exhibit longer learning cycles that arise due to tighter resource constraints. Developing countries are generally considered to be in a disadvantaged position in the global system of knowledge production. They are understood as peripheral from the perspective of industrialized countries. However, some developing countries, especially in East Asia, have managed to enter the stage of original knowledge producers. However, this recent upgrading capacity on the global scale reveals growing internal hierarchies of the national science and innovation systems, creating a division between domestic centers and peripheries.

Biotech in China develops largely through state-sponsored programs (Zhang et al., 2011). This is in line with other aggressive government funding strategies intended to promote science leapfrogging and integration into global systems of knowledge production, such as in the field of nanotechnology (Appelbaum et al., 2011). These state-centered economic sub-systems are necessary to prevent an under-investment in the private sector at immature stages of development. Just like the interdependence of hierarchies on the global scale, "the Chinese model represents a complex mixture of centralized and decentralized elements" (Appelbaum et al., 2011: 309).

These structures provoke the use of terms developed in world systems theory, polarization theory, or dependencia theory. And indeed, some processes that shape hierarchy in science systems may be similar to those shaping economic systems. However, it seems more promising to make use of the terms which describe hierarchical and spatial patterns while looking at their development through the lens of network theory. This approach promises to provide some new insight into world systems and their structures and dynamics.

The obvious state-centricity of biotech funding in China, its potential hierarchical interdependence of national sub-systems and the strategic impact of basic research on the economic valorization all make this industry a suitable case for investigating network structures and development dynamics in a competition-intensive regional setup in China. This article aims to address the conceptual as well as the empirical scarcity of spatial aspects and hierarchies in networks, as suggested for further research by Chase-Dunn and Jorgenson (2003: 13) or Frenken et al. (2009).

The remainder of the article is structured as follows: after a short review of contemporary approaches to spatial hierarchical systems and networks of players in section 2, the subsequent section 3 will briefly introduce the novel approach to measuring the regional capacity of mediating flows in networks and introduce the data. Section 4 analyzes the evolution of the biotech science system in China based on the mediating power of cities and provinces, and will track the changes over time. Section 5 discusses the results and presents consequences and conclusions for the debate on hierarchies in socio-economic systems.

## 2 Theoretical conceptions and prerequisites in the Chinese biotech system

Center-periphery patterns have attracted the interest of researchers from of a broad field of disciplines who look at these patterns from different angles. A most prominent example is world systems theory, and the view that the existence of center, semi-periphery and periphery is continuously reinforced by the logic of the capitalist system. But center and periphery may also be understood as a feature of networks (insiders and outsiders). They may be a purely territorial description, for example of wealth distribution, or they may depict power relations and governance structures. An interesting turn of events has renewed scholarly interest in world systems and spatial center-periphery patterns. The work of Gereffi and others on global commodity chains (GCC) and global production systems or global production networks (GPN) has introduced a player-specific view and made clear how center-periphery is reproduced at the micro-level of products and interaction between companies. Many scholars are "using GCC as part of a world-systemic method as comparative analysis for nested and over time comparisons" (Ciccantell and Smith, 2009: 364). Moreover, this approach has made it possible to take a closer look at upgrading, i.e. the purposive change of an inferior position held by an individual player, a company, a region, or a nation. The ability to upgrade is set in the *seedbed of change* in semi-peripheral places (Chase-Dunn and Hall 1997). Analogies of this conceptual and analytic approach can be found in science and technology (S&T) systems, with respect to territoriality, network structures, hierarchies and upgrading. However, the causes underlying these processes differ from what we see in the economy. Exploring this point, however, would go beyond the scope of this paper. Those interested may consult Gereffi (1999) for more on this issue.

World systems theory and the GCC approach have their conceptual equivalents in scientific systems, specific *epistemic communities* and collaborative research (Weisberg and Muldoon 2009). An increasing specialization of knowledge may induce cultural fragmentation and the creation of new barriers (cf. Bonikowski, 2010). However, an increasing international division of scientific processes can also be expected to help transcend hierarchical boundaries and subsequently integrate peripheral science systems into the global core (cf. Hennemann, 2010).

Compared to the original world system perspective, there is an interesting difference in knowledge systems and their epistemic communities: there is not necessarily a domination of core players over peripheral players in a system of dividing knowledge production at different steps. Cumulative learning effects are suited to facilitating an upgrade of peripheral players. The special character of the good *knowledge* (e.g. non-rivalry) makes knowledge systems different from other systems such as industrial production chains. In knowledge exchanges, all participants are likely to expand their individual knowledge pools (cf. Grossman and Helpman, 1991). Collaboration in the case of knowledge production is therefore different from other commodity exchanges, since knowledge has very different properties of valorization, i.e. knowledge usually gains in systemic value with each transfer act. This may lead to the confusing situation that under certain conditions, peripheral players will benefit (=learn, accumulate knowledge) much more from collaboration than the partner at the semi-periphery or in the core. Certainly, the gradient of individual knowledge stocks should not be too large in order to combine thoughts for a systemic benefit (cf. Pack and Saggi, 1997). But overall, integrating into global knowledge networks can help to break up spatially bound world systems, as discussed by Chase-Dunn and Jorgenson (2003: 4), and subsequently to improve the position of (sub-national) peripheries – not only in the narrow sense of the S&T system, but also in the broader sense of the industry as a whole.

Therefore, it is mainly the political and economic complexity of the science and technology system (i.e. the knowledge pool gradients), which *differentiates* the global system of knowledge, rather than producing boundaries based on true *hierarchical* dependencies with one part of the system dominating the other(s) (cf. Chase-Dunn and Hall, 1997). Temporary differentiation can persist, but dependencies are limited in time because of unintended spillovers of knowledge, especially of scientific knowledge. In a phase of persisting knowledge gradients, the resulting differentiation may produce power imbalances that are present in core/(semi-)/periphery interactions (cf. Chase-Dunn and Jorgenson, 2003: 9). If power constitutes the differentiation process, the simple micro rule of *preferential attachment* (PA) can explain the attraction and the relations. In the PA model, newly entering players prefer to link to strong players that further increase their own importance (Barabási and Albert, 1999). This has been referred to as the *Matthew effect* and describes the effect of amplification of initial states in both positive and negative directions in social interaction (Merton, 1995). In such cases, power develops from initial advantages in resources and potential benefits that important players can provide to newcomers in the system. As opposed to other environments, however, there are few knowledge monopolies. This increases the bargaining power of newcomers, which can choose from more than one potential partner. Behind the linking process in knowledge creation, there is a complex mechanism in which it is not only important how frequent the existing connections of a chosen partner are, this partner also has to accept the connection offer. This reciprocal interplay can be described by utility and cost functions (Goyal, 2007). In the case of peripheral players from developing countries in knowledge networks, this attractiveness can come from specific market environments (e.g. favorable laws, huge customer potential). Power in this understanding of knowledge systems is, therefore, not necessarily characterized by domination and exploitation, because in knowledge creation processes, both partners will usually experience positive effects through learning.

An attractive attachment position can be accomplished not only by (random) initial advantages (i.e. the preferential attachment mechanism in a narrow sense), but an efficient utilization of resources and dexterous linking to complementary information may also result in greater attractiveness than other participants in the network. Such preferred hubs in knowledge networks can be considered temporarily powerful with respect to knowledge transmission. However, this situation may not last for long, since the complexity of newly entering players can instantly rearrange the whole topology of the system. In science systems, therefore, initial asymmetries neither necessarily lead towards a crowding out of the periphery, nor do they produce manifest structures of information dominating nodes.

Consequently, recent contributions acknowledge an upward mobility potential at the semi-periphery and consider China to be taking a lead for other developing countries in the future in this respect (Mann, 2010: 179). This is not only assumed for pure economic activities, but also expected for increasingly relevant contributions from the Chinese science system in some highly state-sponsored key science fields (cf. Zhou and Leydesdorff, 2006).

It becomes clear that the pure territorial view of political and economic black boxes is too narrow for the explanation of knowledge systems and their consequences for the socio-economic development. The spatial sphere has to be expanded with the integration of a network perspective of flows, since network centrality measurements are an important indicator for the translation of world systemic patterns (Chase-Dunn and Jorgenson 2003,12).

*Concept of regional centrality in a networked knowledge system*

In a combined perspective of networks and space, Liefner and Hennemann (2011) hypothesize that agglomerations may show distinctive features relative to their network embedding and their capability to broker knowledge between places. In their typology, regional lock-ins and a low capacity for knowledge brokerage may lead to long-term economic downturn even in large and heterogeneous agglomerations. In turn, the ability to broker knowledge in the current system state may enable smaller, sometimes economically backward regions to raise their economic power in the future. Processes of integration and disintegration have to be acknowledged, which means not just following a structural description of categorical power of places (cf. Robinson, 2002: 548), but also taking into account the influence of relations and strategic positions in networks that enable controlling flows (cf. Burt, 1995). Placing importance on *spatial flows* is not in contrast to the centrality of places (e.g. in *Christaller's* sense), but encompasses this static view and adds a dynamic perspective of *flows and structural development* (cf. Taylor et al., 2010). An efficient attachment to others in networks will almost automatically enable players with scarce resources to improve their situation significantly. In early stages of development, smart public spending may, therefore, facilitate the integration of players and regions into sophisticated global systems.

Particularly in developing economies, there are only some regions capable enough to adapt to these dynamics and to attach to the centers of global knowledge production networks. In an initial evolutionary stage, they are essentially brokering between the global centers and the sub-national periphery. In China, the strong division into a two-tier system is supported by huge central funding programs (e.g. projects "211", "985") that are open only to selected organizations. The central government aims to develop a competitive research system on a worldwide scale with this sub-national border-creating funding scheme. It is clearly an explicit growth-related policy instrument, leaving many other public research organizations in a disfavored situation.

However, we argue that many provinces in China possess features of the semi-periphery in the global system of knowledge production, and, more importantly, that the ability to broker knowledge can change during the evolution of the network, i.e. peripheral regions can improve their relevance for the network flows over very short periods of time. Due to the strong interaction between the science sector and the economic sector in technological paradigms such as biotechnology, the ability to attach to global science networks will have a direct and indirect effect on the economic development potential (Gertler and Levitte, 2005; Herring, 2007). Therefore, universities and public research organizations are assumed to play a significant role in early stages of product life cycles.

*Science-based industries and their developmental impact in China*

In China, most biotechnology-related research offers great opportunities to overcome resource scarcity. This part of the science system may thus be considered sensitive for the socio-economic development of China as an enabling technology for many other sciences, and a *window of opportunity* for the economies (cf. Niosi and Reid, 2007).

Examples of the deep socio-economic impact of biotech for developing countries include productivity gains in harvest due to higher resistance to rodents, or cultivation expansion to formerly unsuitable areas with respect to natural conditions. Productivity gains in rural areas were the initial boost for the economic development of China in the 1980s/1990s, and are still of interest because the population is still growing and the arable land is increasingly scarce. In the case of research in and production of genetically modified crops, China is among the world's leading countries, also due to its strong public research sector (Herring, 2007). This strategic and policy-driven development of the agricultural system is seen as a major step for securing food and reducing poverty in developing countries (Adenle, 2011). Population health is another area in which biotechnological research will contribute to an overall improvement of the living conditions. Inexpensive pharmaceuticals and treatments are needed especially in poor, non-urban areas of all provinces in China with a high growth potential (Frew et al., 2008). Currently, mainland China is a market worth USD 8 billion p.a. in biotech, with growth rates of 20% each year for the last 5 years (Datamonitor, 2010). More than 90% is generated with pharmaceuticals, while another 6% of the market is related to crops and agriculture. Compared to other sectors, this market is rather small, but with outstanding growth potential.

The need for collaboration is commonly explained with resource-related, strategic arguments that also hold for collaboration in the biotech sector. The lack of their own resources in the production of new knowledge urges firms and universities alike to work together, sharing thoughts and apparatuses. This is especially true for the highly specialized nature of scientific and product-related knowledge in biotech (Thorsteinsdóttir et al., 2010). And it is not only the developed system parts which have something to offer in collaboration activity.

China has its own traditional way of dealing with epidemics and related issues that can contribute towards new ways of problem solving (Thorsteinsdóttir et al., 2011). Hennemann et al. (2011) showed that collaboration patterns for global epidemic research and the involvement in H5N1-related research are significantly different than for other scientific sub systems (e.g. nanotech), involving several countries from the southern hemisphere. Therefore, this capacity-building argument goes hand in hand with economic development in developing countries, access to materials for both sides, access to expertise and technologies. With this advancement, China is found to play a mediating role between the developed systems and least developed countries – collaboration between these mediators and the core countries is increasingly collaboration among equal partners (Thorsteinsdóttir et al., 2011).

Involvement in these systems can enhance the capacity to innovate, for example through the assimilation of new ideas from the global knowledge system through interconnection to those places. Chang (2003: 434) shows for Taiwan and the UK that those biotech firms which are involved in government programs are more likely to create successful innovations. This success is mainly based on the accessibility of leading scientists who secure immediate access to high-quality global networks and the delivery of complementary information to the collaborating firms. For mainland China, there is still a significant impact of public research organizations on the knowledge production and dissemination for the Chinese system. Patents are usually still held by universities, key laboratories or other public institutions, rather than by businesses (Chen and Guan, 2011). Additionally, the collaboration activity of Chinese organizations with players from developed countries is limited when compared to other developing countries such as India (cf. Melon et al., 2010). However, beyond this international perspective, there is a remarkable blind spot in the current debate when it comes to the internal structures and dynamics of biotech systems, especially in China. Therefore, our interest is mainly focused on the intra-country pattern and the regional capacity to mediate knowledge flows in that system.

**3 Methods and Data**

Science and knowledge production are likely to be the prerequisite for economic and societal advancement as explained in the previous section. However, for measuring activities in world systems and center/periphery interrelations, it is necessary to evaluate micro-foundations of individual interactions, i.e. to assess the quantities and qualities of player *relations* (cf. Chase-Dunn and Jorgenson, 2003). Developments and activities in science production are frequently evaluated by simple paper or citation counting that is done at a territorial aggregate level (e.g. countries). This equates centers of activity with those regions that produce the most scientific papers. As explained above, there are regions that are limited in their capacities, but are very effectively attached to central players in the knowledge production system. If we seek to evaluate the efficiency of knowledge production and the ability to influence the knowledge dissemination that serves as a baseline for economic development, simple counting methods are insufficient, because they miss those players which produce little, but in a very influential way. Therefore, player-based relations are best reproduced by graphs or networks, an approach that is frequently used in sociology and connected fields. We do not employ citation-related indicators, because they have proven to be inaccurate in representing strong relations between system players. However, using co-authorships to produce networks has its own difficulties, because in comparisons that involve different development stages of networks over time, the complexity of the systemic interactions makes assessments based on descriptive network measurements implausible. This is because networks are largely $n,k$-dependent, i.e. network properties such as graph-level indices or node-based measurements are generally incomparable between networks of different sizes without adjustments (cf. van Wijk et al., 2010, Hennemann et al., 2011b). To account for these issues, we use a randomizing approach based on edge-swapping that was first proposed by Maslov and Sneppen (2002), and technically enhanced with a Markov Chain realization for connected graphs by Gkantsidis et al. (2003). For statistical testing and the calculation of parameter estimates, we use a bootstrapping approach, which is explained in detail in Hennemann (2011). The calculations were made using the package NetworkX for python (http://networkx.lanl.gov/).

*The data*

In contrast to other empirical studies that use case studies, patents or participation in sponsored research consortia to assess the knowledge-producing system, we propose two readily available publication databases (SCI-Expanded and China's Chongqing VIP) and use *co-authorships* to build collaboration networks.

Recent research has shown that the assessment of regional network phenomena in dynamic scientific fields is especially difficult to handle in the context of developing countries. International publication data from sources such as the *ISI Web of Knowledge$^{TM}$* overemphasize the globally active organizations, whereas local domestic sources such as the *Chongqing VIP database* virtually neglect international collaboration in science (Hennemann et al., 2011b). To reduce the regional bias, the two sources are combined to produce a comprehensive overview of the evolution of this increasingly important scientific field for China.

Table 1: Size of the combined SCI-E/VIP networks and their largest components H[0]

|  | 2001 | 2002 | 2003 | 2004 | 2005 | 2006 | 2007 | 2008 | 2009 |
|---|---|---|---|---|---|---|---|---|---|
| Total number of Nodes | 351 | 531 | 649 | 852 | 1,108 | 1,390 | 1,690 | 2,023 | 2,272 |
| Total number of Edges | 332 | 567 | 777 | 1,116 | 1,693 | 2,376 | 3,270 | 4,061 | 4,574 |
| Nodes in H[0] | 170 | 313 | 431 | 623 | 918 | 1,174 | 1,459 | 1,764 | 1,961 |
| Edges in H[0] | 208 | 428 | 633 | 968 | 1,570 | 2,233 | 3,110 | 3,894 | 4,383 |

Note: in recent years (2007-9) almost half of the nodes were located in coastal provinces and half of them were located in interior provinces (assumed periphery)

The international data come from ThomsonReuters ISI Web of Knowledge[TM] SCI-E index and contain only those articles that involve at least one co-author with a Chinese affiliation[3]. The local data were collected from the domestic Chinese database Chongqing VIP, using the same selection criteria in a variant, because the database cannot be searched, but rather has to be compiled manually. The combined dataset ISIVIP consists of 9,192 raw articles, with approximately half coming from each of the original sources in the years 2001 to 2009. These papers were processed to construct networks of co-authorships for each year at the organization level in such a way that all contributors of a paper were connected to one another. Connections between two organizations were maintained for two successive years, even if no additional paper was produced to account for the continuing awareness of individuals for each others' work ("window of cooperation"). Collaboration within the same organization was not included, leading to a true inter-organization collaboration network. Table 1 shows the size of the networks as well as the size of the largest connected component H[0] in each year. All calculations are applied to the largest connected component.

The calculations in the following section frequently use China's 30 provinces as the scale at which a distinction between local and inter-regional collaboration is made. China has 30 provinces, with sizes ranging from several thousand to more than a million square kilometers, and from less than ten million to more than 100 million inhabitants. Most importantly, the provincial science and technology systems differ greatly with respect to input and output. Each provincial system, however, is usually dominated by a few leading organizations that receive preferential funding. Hence, domestic collaboration in the network can be intra-provincial or inter-provincial.

**4 Analysis and Results**

*General structure of the spatial network*

The biotech science network grows moderately in the first half of the period under investigation and then almost doubles its pace. The mean geodesic path length decreases slightly over time, as does the average clustering. Considering the huge increase in size, the network increasingly shows small-world properties, i.e. short average path length *and* reasonable clustering. Interestingly, although most social networks have positive degree correlations (cf. Newman and Park, 2003) the biotech collaboration network here shows a negative degree correlation that is only slightly turning towards an uncorrelated state. Warren et al. (2002) attribute this effect to the special structure in geographic networks that is induced by spatial clustering in agglomerations; a result that has been confirmed by Hennemann et al. (2011a) for global science activities.

This clustering is reflected in the general spatial pattern. Firstly, the collaboration probability *p(d)* is negatively related to the spatial distance *d* between the collaborating parties for all the years from 2001 to 2009[4]. Secondly, this

---

[3] The title, keywords and abstracts had to match the following term: "CU=China AND (TS=((Cell OR ENZYME OR APPLIED) SAME Bio*) OR TS=Bionic)"

[4] We calculate the conditional probability in logarithmic bins with the fraction of nodes that are linked from all nodes in the network at the given distance. For a detailed description of the

distance dependence is more significant for collaboration within the same province than across Chinese provinces. Inter-province collaboration shows features of random partner selection. There is an increasing collaboration probability with distance for collaboration that involves foreign partners, i.e. long-distance collaboration is more likely to occur than collaboration with partners in neighboring Asian countries (Fig. 1). This relation is very stable over time and shows systematic deviations from the randomized null model, i.e. the findings are significantly different from pure chance, given the empirical network topology.

This result is robust against simple coast/interior effects, except that intra-coastal collaboration is slightly more likely than intra-interior collaboration. The results have also been controlled for the effect that the municipalities Beijing and Shanghai may cause. Especially in early stages of the development, a large fraction of the activity is focused on players from these two provinces. However, territorial provinces show similar distance dependency patterns to the city provinces. The higher relevance of the shorter-distance collaborations is a quantitative measurement and is usually due to the need for intensive face-to-face contact in complex negotiation during highly unstructured knowledge creation processes. Here, network proximity and spatial proximity overlap to a large degree, which is surprising, because intra-organizational collaboration was excluded from the data.

Social network theory suggests that it is often not the local connection that matters, but those connections between players that are spatially distant from one another (cf. Granovetter 1973;

calculation procedure see Hennemann et al. (2011a)

Watts and Strogatz, 1998). Therefore, the strictly local clustering of collaboration activity may not be the most important with respect to the ability to influence knowledge production. In crossing territorial borders (e.g. provincial, national), heterogeneous knowledge pools which one cannot find around the corner may become accessible, making these connections more important, because the outcome is of higher value (cf. Guimera et al., 2005). This is supported by the empirical data of Jones et al. (2008) for inter-organizational collaboration compared to intra-organizational collaboration in science. Jones and colleagues used citations to assess the quality of academic work.

Generally, the quality or the value of collaborations is difficult to assess. Bibliometric analyses suggest using citations (scientific articles, also patents), but this has some shortcomings, especially for very young science fields, as there is a time delay in citations. Besides this practical issue, there is also a conceptual problem present. Is it the quality of scientific work which is reflected in high numbers of citations, or is it simply a matter of prominence? - i.e. frequently cited papers have a greater chance of receiving further citations compared to publications with few citations (again, a *Matthew effect*). Moreover, citation is a loose and arbitrary construct that may also be used on the basis of strategies and not based on content. Here, we assess the quality with the edge flows (=loads) in the collaboration network, which assumes that those connections that are central (=crucial) in the science field are most important for the network functioning with respect to flows. The load of an edge $e$ is given by the sum of the fraction of all-pairs shortest paths that pass through edge $e$ (cf. Brandes, 2008).

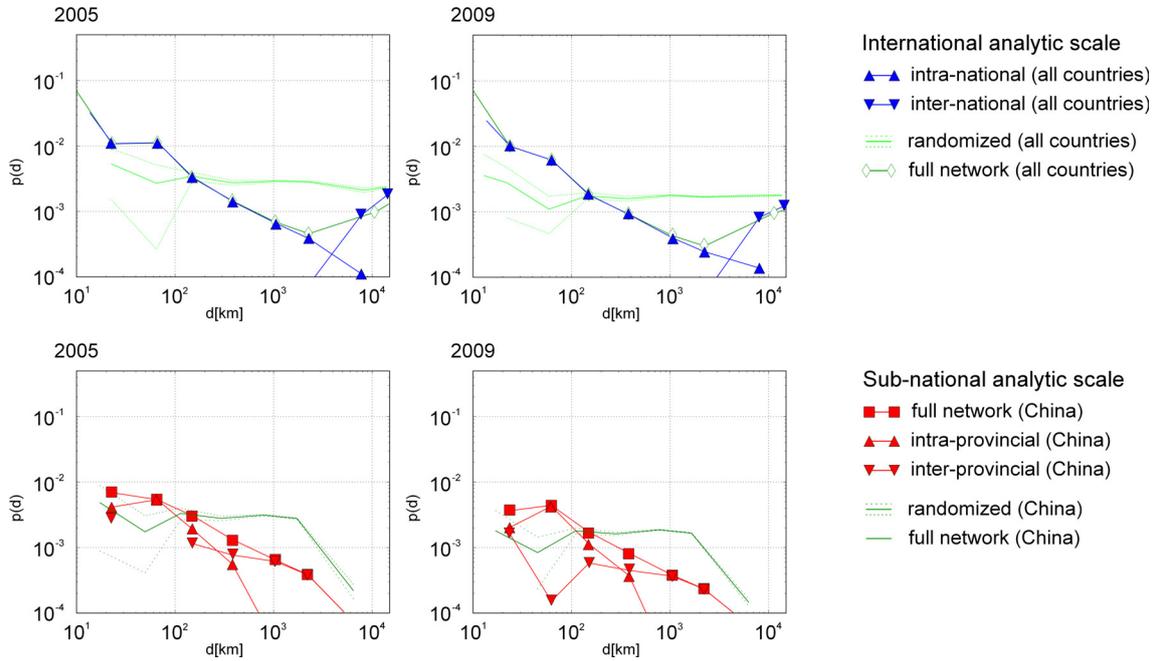

Figure 1: Collaboration probability in dependence of the Euclidian distance d for the years 2005 and 2009, own calculation

Note: the similarity is present in all other years as well

Figure 2 shows the empirical edge loads for different connection scales as well as the randomized expectation that can be derived from the null models.

Generally, the empirical loads of edges that connect nodes within the same province are significantly lower (Fig. 2, left panel) than the loads of edges that cross-connect provinces (center panel). Local edge loads are within 2 standard deviations (SD), therefore not significantly different from the expected values that are based on the randomized network. In recent years, this has changed slightly towards a load that is *lower* than expected. The inter-province edge loads (Fig. 2, center panel) are significantly higher than those in the corresponding randomized networks, indicating an important function of collaboration that crosses province borders and levels the regionally specific knowledge pools. Inter-province linkages that cross boundaries fulfill the function of stabilizers, i.e. serving as a backbone for the network. Beyond that, these long-range connections will provide access to cognitively distant knowledge, avoiding local lock-ins.

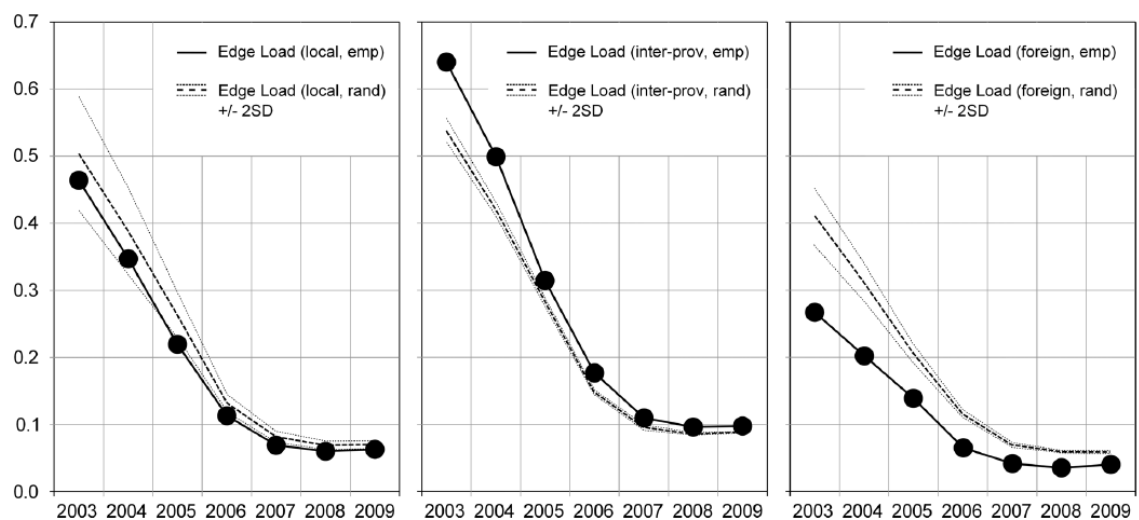

Figure 2: Average edge loads for local (intra-province), inter-province, and international linkages compared to randomized versions of the respective networks.

The loads of edges that involve foreign organization are systematically lower than expected (Fig. 2, right panel). As already mentioned, this can mainly be attributed to the network construction that favors Chinese over foreign organizations to be selected into the network.

Figure 3 shows normalized[5] edge load values that can be compared across the networks over time. Local collaboration (green line, square), although of higher probability (see above), is less important for the overall knowledge flow in the network, but has been gaining relevance recently. However, the magnitude of the edge load is lower than expected by chance. In contrast to this, inter-province collaboration is much more relevant compared to local collaboration with respect to knowledge flows in the network. Recently, the relevance of inter-province collaboration has been decreasing, while the collaboration with foreign involvement has regained importance, with North-American cities in particular becoming more relevant. This is interesting because there is a bias arising from the data selection that inflates Chinese organizations over foreign ones.

---

[5]The normalization was carried out by dividing the empirical edge load by the randomized value. This calculates the average edge load with respect to an ideal uncorrelated graph possessing the same basic properties. As a result, the network parameter $P_{rand}$ is directly comparable over time.

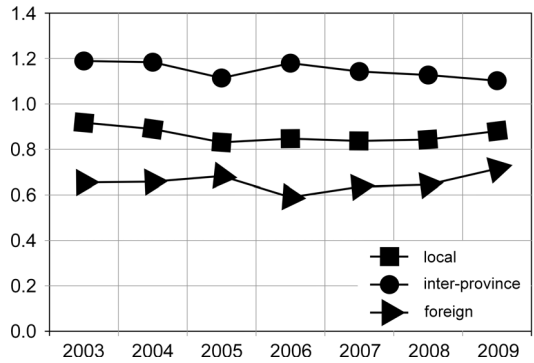

Figure 3: Normalized average edge loads for local, inter-provincial and foreign connections, own calculations.

*Regional capacity to influence the knowledge flows of the entire system*

Based on this strong quantitative evidence of spatially clustered networks that are different from long-range connections with respect to quality improvements for the network, the most important location in China for the knowledge production and the integration into the world system of knowledge can be identified. Our main indicator to track the knowledge flows in the empirical networks is a weighted spatial aggregate of the betweenness centrality of nodes $C_B(v)$. It uses the k-core values of the nodes as a weighting to acknowledge the dramatic difference of central compared to peripheral nodes, *ceteris paribus*, to influence the network topology and flows (Kitsak et al., 2010). Therefore, nodes that are central in the network are given a greater weight in the aggregation procedure.

Figure 4 shows the results of the development of Chinese cities for the years 2001 to 2009. One striking feature is the decreasing relevance of foreign nodes from neighboring countries such as Japan. Recently, North American cities have become more important for the Chinese biotech network. However, they are not displayed in the maps. The second important element is the decreasing relevance of the coastal cities, turning the network into a less hierarchical one, i.e. not dominated by the leading organizations in Beijing and Shanghai. The southern cities, including Hong Kong, were the prominent knowledge disseminators in 2002, but have been of little influence in recent years. The coastal cities around the Yangtze River Delta (e.g. Shanghai, Nanjing, Hangzhou) and the Bohei Region (e.g. Beijing, Tianjin) were most relevant for the network in the years 2003 to 2005, and have become less relevant in recent years.

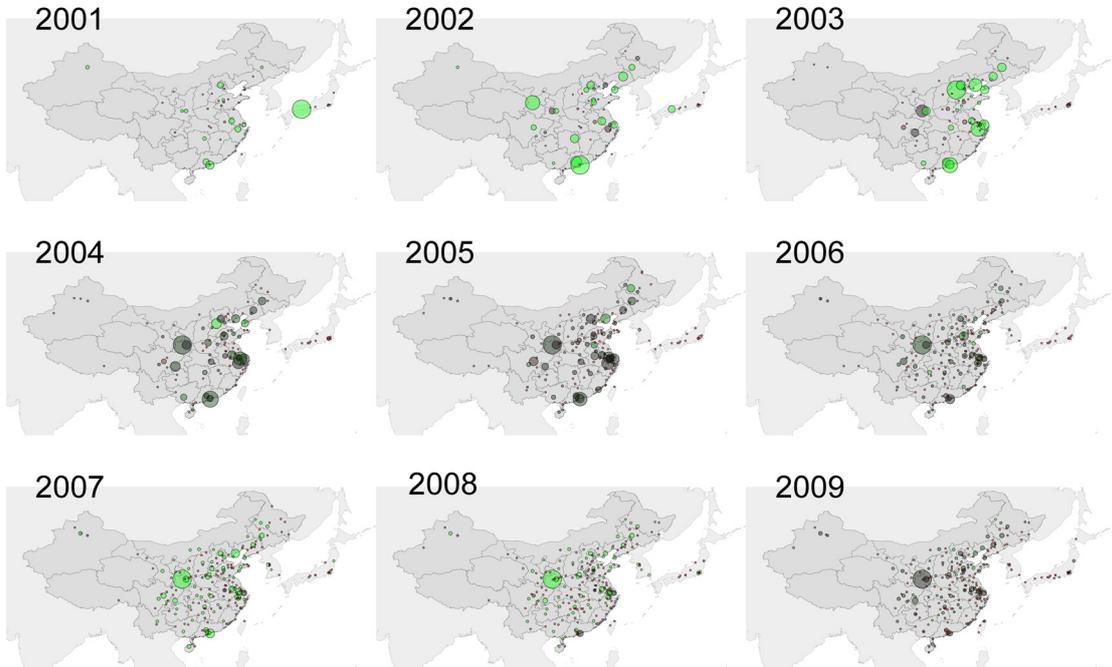

Figure 3: City network performance development in the biotech science system from 2001 to 2009

Note: circle size indicates the weighted aggregate betweenness of city nodes, the efficiency of a city is defined by its empirical betweenness C_B (v) compared to the estimate from the null model $C_B(v)_{rand}$

Green:   over-performing, efficient city (emp/rand >= 1)
Red:     under-performing, inefficient city (emp/rand <1)
Grey:    efficiency according to expectation (emp/rand ~1)

However, there are three other areas in the interior that deserve interest: Sichuan, Chongqing and especially Shaanxi. These are all provinces that are in the center of the agricultural production system, and the research organizations are consequently pushed into these research streams, not only by central government initiatives, but also by provincial funding. Shaanxi in particular, and the provincial capital city of Xi'an, has been a very important and stabilizing city node in the science network in the years from 2002/3 until today.

From an evolutionary perspective, the performance in the past may direct the performance in the future. In the case of Xi'an, the presence and efficient use of direct connections to others in early stages of the development can be seen as being predictive for the great importance recently (cf. tab. 2). The cross-correlation between $C_B(v)$ and $C_B(v)_{rand}$ suggests a time delay of several years.

Interestingly, early poor performers show a tendency to become efficient later, while the efficient cities in early stages occupy strong positions after several years. However, the strong cities face difficulties maintaining their efficiency. This suggests that players are attempting to stimulate their relevance consciously by attaching themselves to promising partners, and that the early phase of the network evolution is decisive for the centrality in later stages. However, the time series is too short to prove robustness of these tendencies.

| | Empirical Betweenneess in year | | | | | | | |
|---|---|---|---|---|---|---|---|---|
| Efficiency/Performance (normalized betweenness) at time | | 2003 | 2004 | 2005 | 2006 | 2007 | 2008 | 2009 |
| | t+6 | -0,59 | | | | | | |
| | t+5 | -0,45 | -0,62 | | | | | |
| | t+4 | -0,14 | -0,45 | -0,68 | | | | |
| | t+3 | -0,19 | -0,20 | -0,54 | -0,65 | | | |
| | t+2 | 0,14 | -0,21 | -0,34 | -0,52 | -0,63 | | |
| | t+1 | 0,40 | 0,16 | -0,22 | -0,33 | -0,48 | -0,58 | |
| | t | 0,69 | 0,53 | 0,16 | -0,14 | -0,21 | -0,40 | -0,46 |
| | t-1 | | 0,64 | 0,32 | 0,05 | -0,16 | -0,25 | -0,37 |
| | t-2 | | | 0,54 | 0,26 | 0,06 | -0,11 | -0,21 |
| | t-3 | | | | 0,44 | 0,35 | 0,10 | -0,06 |
| | t-4 | | | | | 0,49 | 0,29 | 0,11 |
| | t-5 | | | | | | 0,44 | 0,26 |
| | t-6 | | | | | | | 0,46 |

*Regional knowledge circulation*

A conscious stimulation of one's own importance in a network is likely to be achieved through a wise linking to the right nodes in a network. But which player is the right one? We address this question by analyzing the dense areas of the network and the interaction between these dense areas. Newman (2006) suggested a technique for the detection of communities in complex networks. The proposed *modularity* is used to find players that are intensively inter-connected to one another. The properties of the partition (e.g. regional origin of the organizations) can be interpreted and described to find sub-communities that circulate knowledge. Regions that are prominently present in different sub-communities are a prospective broker of knowledge between different regions.

In appendix 1, an example of the modularity-based network clustering is provided for the year 2009. For clarity reasons, not all years are presented here. There are 21 modules in the network (labeled A to U in appendix 1), of which only 10 show significant activity (these represent approximately 95% of all edges). Two tendencies can be derived from the module analysis: firstly, coastal provinces are the main utilizer of foreign collaboration. Shanghai, Beijing, Jiangsu, Guangdong, Zhejiang and to a lesser extent Shaanxi, Sichuan, and Henan as interior provinces are strongly connected to foreign players (modules A, B, C, E, F, G, H, I, appendix 1). Secondly, Module D interacts comparatively strongly with foreign-dominated modules on the second level, i.e. indirectly. These connections involve foreign nodes and nodes from Beijing and Shanghai. Therefore, it can be hypothesized that the Shaanxi province has the function of a backbone that mediates between foreign nodes and primary coastal nodes. This result confirms the strong capacity of Xi'an (see above). However, it may well be that this mediation is only occurring among those organizations that are very present in the network, leaving out the weaker ones.

*Summary of analytical results*

These results point towards a complex multi-scale system of science in China's biotech research, where network structures strongly interfere with spatial structures:

A) patterns observed:

- The bulk of collaboration is a localized activity, and most linkages do not cross province boundaries. This may be counter-productive for an effective integration into the world knowledge system, because each of the provinces has to seek international contacts individually.
- The relatively few connections that span provincial borders are more important.

B) dynamics:
- The spatial structure is shifting from a comparatively strong foreign influence through a significant importance of Hong Kong, Shanghai and Beijing towards a general importance of coastal cities. Hong Kong in particular is becoming marginalized over time. This dynamic change reflects China's integration into the world science system (reduced foreign dominance and rise of the leading national centers) which goes hand-in-hand with the emergence of a national center-periphery pattern. Later, the domestic structure becomes more balanced (reduced dominance of the leading national centers).
- Some interior cities in Shaanxi, Chongqing and Sichuan are becoming increasingly important mediators for the whole system, although foreign activity is still related to collaboration with coastal players.

Consequently, the knowledge circulation in the system is becoming relatively heterogeneous and unbiased over time (leveling/democratizing function of knowledge dissemination).

*5 Discussion and Conclusion*

The approach presented offers novel insight into the organization of player relations and their dynamics that lead to rearrangements in center-periphery structures on the global scale and, more importantly, it delivers a detailed view of internal changes in the (semi-)periphery in globalized science systems, which are the assumed *seedbeds of change*. We used a player-centered network approach that has proven to be a solid method of evaluating the structures and dynamics in center-periphery-driven systems.

The center-periphery pattern in world science systems exists, even for large and dynamic science producers such as China. However, the sub-national pattern did not show strong persistence, but remains open for changes in the seedbed in semi-peripheral places. This upgrading potential in knowledge-based industries has often been overlooked in the recent player-centered discussions that comment on hierarchies in international networks, such as GCC-type concepts. We propose adding a perspective of knowledge networks into these concepts to strengthen their capability to explain peripheral upgrading in increasingly science-based industries. Upgrading is the key force at play that enables efficient utilizers of knowledge to integrate quickly into flows of knowledge. Moreover, the special character of the good knowledge as well as the learning differentials can help to control the flows very quickly, as in the case of the Shaanxi province. Concerning the sub-national patterns, there is no continuous domination of the coastal semi-periphery over the interior periphery, because there is a normalizing (marginalizing?) effect of the coastal influence on the knowledge flows over time. The spatial structure is shifting away from a comparatively strong foreign influence through the usual mediators Hong Kong, Beijing and Shanghai towards a general importance of coastal cities. Even some interior regions in Shaanxi, Chongqing and Sichuan are becoming increasingly important mediators for the whole system, although most foreign activity is still related to collaboration with coastal players.

However, with respect to sub-national center-periphery structures in China, the east still contributes much more to the internationalization of the Chinese biotech network compared to the western provinces – but the knowledge is increasingly being influenced by players from the interior. Similarly, China is taking on an influential position in the mediating process between the global centers and other developing regions outside China. This may have interesting consequences for the global north-south balances.

The quick rearrangement of the spatial structure can be interpreted as change in the temporary domination of cores over the periphery. As such, the power-asymmetry is likely to be induced by temporarily imbalanced regional knowledge pools, rather than by persistent core-periphery dominations that are caused by hierarchies. Moreover, the function of the Shaanxi province as a mediator between foreign knowledge and the interior as well as the coastal provinces and the interior shows that there are multiple relations between players possessing different abilities to absorb, produce and disseminate biotech-related knowledge. This makes the complex system of knowledge production in China's biotech science a system of high differentiation rather than a system of hierarchy. Overall, the knowledge circulation in the system is becoming relatively heterogeneous and unbiased over time (leveling/democratizing function of knowledge dissemination).

The knowledge-centered approach to socio-economic development in science-driven industries can explain fast-lane development in some areas of biotech. For example, in the field of traditional herbal medicine research and product development, China is one of the global market leaders (Thorsteinsdóttir et al., 2011). Recently, the China-based Beijing Genomics Institute, an academic spin-off company in Shenzhen, was successful in deciphering the genetic code of the EHEC virus that caused an epidemic threat to wide areas of Europe, including Germany, in early 2011. There, a sophisticated DNA decoding technology was used to process the cell material (Enserink, 2011). This expertise is increasingly being used for capacity building in other developing countries, making China a perfect mediator of knowledge between the global centers of biotech and backward regions in Africa. Such "south-south" collaborations are seen as an alternative to classic center-periphery collaboration (cf. Thorsteinsdóttir et al., 2011).

The greater complexity of activity in the core of the semi-periphery seems to be advantageous in early stages of the development. However, our results have shown that early leaders have difficulties retaining their superior position in the network, and that newcomers can indeed take a lead after very short time periods. The system analyzed is basically a science system, but the close interrelation between science and business valorization in biotech can be assumed to lead to a rearrangement of the development perspectives as well. However, this is only an assumed indirect effect, although many firms and private organizations are included in our data.

With increasing differentiation of global knowledge systems, the classical notions of product-specific networks, as in the case of GCC, do not hold true when largely knowledge-driven sectors become more relevant for the socio-economic development. From a theoretical perspective, science-induced global epistemic communities can help to overcome the deficit in explaining the rapid catch-up process of China in some academic and industrial sectors. The integration into narrowly focused knowledge networks (i.e. epistemic communities) offers opportunities for marginalized players to participate and grow in importance. Knowledge networks thus potentially democratize increasingly complex world systems, because the accumulation of knowledge is becoming a strategic resource for players in developing economies.

The policy implications of our results are directly related to this explanatory power of knowledge networks in China. The strong political initiative in infant science-based industries, such as in the biotechnology case, can be seen in many other sectors in China at the early stages of development. Interestingly, most of these initiatives can be assumed to play a significant role in regionally differentiated economic development. Therefore, this Chinese way of "sculpturing" the seedbed of change at the semi-periphery may serve as a role model for other large developing economies such as Brazil, Russia and India. A major obstacle for such a transfer is the difficulty of selecting promising players that may enjoy the benefits of early network attachment. This selection process will require some further investigation.

However, this study is based on biotech, which may be a special case that cannot be generalized due to its strong agricultural focus that is very country-specific with respect to the spatial distribution of the players. Another shortcoming is the lack of pure firm data to indicate the economic activity and development. We did not integrate all global players into the network, which may also create a strong bias in the special situation of the behavior in complex systems. This restriction to those who collaborate with Chinese organizations leads to an underestimation of foreign influence. In future analyses, this fact needs to be addressed by gathering all global publications in the biotech sector for the given time period.

A growing number of network-centered evaluations can be expected in the future, since the field of network science is developing dynamically in both the social (e.g. sociology) and natural sciences (e.g. physics). This will inevitably lead to a greater understanding of center-periphery dynamics and the upgrading of backward regions on the global scale, and subsequently show policy instruments that acknowledge the capacity building nature of knowledge networks.

*Acknowledgements*

The authors are grateful for helpful comments on some of the results that were presented at the International Conference on Urbanization and Development in China 2011, Salt Lake City, UT.

*Funding*

This work was partly supported by the DAAD [travel grant number A/10/00307]; the Alexander-von-Humboldt-Stiftung [grant number TC-Verl. DEU/1131699]; National Science Foundation of China [grant number #40971069]

Appendix A: Module-based network clusters and their interaction with each other (2009)

|   | A | B | C | D | E | F | G | H | I | J | K | L | M | N | O | P | Q | R | S | T | U |
|---|---|---|---|---|---|---|---|---|---|---|---|---|---|---|---|---|---|---|---|---|---|
| A |   | 98 | 96 | 90 | 77 | 32 | 90 | 59 | 98 | 20 | 16 | 7 | 9 | 2 | 2 | 0 | 3 | 1 | 4 | 2 | 0 |
| B | 98 |   | 112 | 55 | 29 | 8 | 44 | 45 | 10 | 36 | 2 | 2 | 4 | 0 | 0 | 0 | 0 | 1 | 3 | 0 | 0 |
| C | 96 | 112 |   | 54 | 28 | 18 | 14 | 40 | 3 | 47 | 1 | 5 | 1 | 5 | 1 | 3 | 0 | 0 | 3 | 1 | 0 |
| D | 90 | 55 | 54 |   | 13 | 21 | 17 | 23 | 7 | 19 | 2 | 0 | 4 | 0 | 0 | 0 | 1 | 0 | 3 | 0 | 0 |
| E | 77 | 29 | 28 | 13 |   | 4 | 15 | 8 | 8 | 9 | 0 | 0 | 1 | 0 | 0 | 4 | 0 | 0 | 0 | 1 | 0 |
| F | 32 | 8 | 18 | 21 | 4 |   | 8 | 7 | 3 | 7 | 1 | 0 | 2 | 0 | 0 | 0 | 1 | 0 | 0 | 0 | 2 |
| G | 90 | 44 | 14 | 17 | 15 | 8 |   | 43 | 3 | 20 | 2 | 0 | 0 | 0 | 0 | 0 | 0 | 0 | 1 | 0 | 0 |
| H | 59 | 45 | 40 | 23 | 8 | 7 | 43 |   | 8 | 16 | 9 | 0 | 4 | 0 | 0 | 0 | 0 | 0 | 4 | 0 | 0 |
| I | 98 | 10 | 3 | 7 | 8 | 3 | 3 | 8 |   | 35 | 0 | 0 | 1 | 0 | 0 | 0 | 3 | 0 | 1 | 0 | 0 |
| J | 20 | 36 | 47 | 19 | 9 | 7 | 20 | 16 | 35 |   | 21 | 1 | 0 | 0 | 0 | 0 | 0 | 0 | 0 | 0 | 0 |
| K | 16 | 2 | 1 | 2 | 0 | 1 | 2 | 9 | 0 | 21 |   | 0 | 0 | 0 | 0 | 0 | 0 | 0 | 0 | 0 | 0 |
| L | 7 | 2 | 5 | 0 | 0 | 0 | 0 | 0 | 0 | 1 | 0 |   | 0 | 0 | 0 | 0 | 0 | 0 | 0 | 0 | 0 |
| M | 9 | 4 | 1 | 4 | 1 | 2 | 0 | 4 | 1 | 0 | 0 | 0 |   | 0 | 0 | 0 | 0 | 0 | 0 | 0 | 0 |
| N | 2 | 0 | 5 | 0 | 0 | 0 | 0 | 0 | 0 | 0 | 0 | 0 | 0 |   | 0 | 0 | 0 | 0 | 0 | 0 | 0 |
| O | 2 | 0 | 1 | 0 | 0 | 0 | 0 | 0 | 0 | 0 | 0 | 0 | 0 | 0 |   | 0 | 0 | 0 | 0 | 0 | 0 |
| P | 0 | 0 | 3 | 0 | 4 | 0 | 0 | 0 | 0 | 0 | 0 | 0 | 0 | 0 | 0 |   | 0 | 0 | 0 | 0 | 0 |
| Q | 3 | 0 | 0 | 1 | 0 | 1 | 0 | 0 | 3 | 0 | 0 | 0 | 0 | 0 | 0 | 0 |   | 0 | 0 | 0 | 0 |
| R | 1 | 1 | 0 | 0 | 0 | 0 | 0 | 0 | 0 | 0 | 0 | 0 | 0 | 0 | 0 | 0 | 0 |   | 0 | 0 | 0 |
| S | 4 | 3 | 3 | 3 | 0 | 0 | 1 | 4 | 1 | 0 | 0 | 0 | 0 | 0 | 0 | 0 | 0 | 0 |   | 0 | 0 |
| T | 2 | 0 | 1 | 0 | 1 | 0 | 0 | 0 | 0 | 0 | 0 | 0 | 0 | 0 | 0 | 0 | 0 | 0 | 0 |   | 0 |
| U | 0 | 0 | 0 | 0 | 0 | 2 | 0 | 0 | 0 | 0 | 0 | 0 | 0 | 0 | 0 | 0 | 0 | 0 | 0 | 0 |   |

A: Foreign, Shanghai
B: Foreign, Beijing
C: Foreign, Beijing, Gansu
D: Beijing, Shaanxi, Hubei, Foreign, Heilongjiang, Henan
E: Foreign, Jiangsu
F: Foreign, Sichuan, Liaoning
G: Foreign, Guangdong
H: Foreign, Zhejiang, Beijing, Guangdong
I: Foreign, Henan
J: Foreign, Beijing, Shandong, Jiangsu, Hunan
K: Hunan, Shandong, Jiangsu
L: Foreign
M: Chongqing, Fujian, Foreign, Sichuan
N: Hong Kong, Foreign
O: Shanghai, Foreign
P: Beijing, Foreign
Q: Foreign, Henan
R: Zhejiang, Jilin
S: Foreign, Hunan
T: Hebei, Qinghai
U: Jiangxi

Note: The network clusters at the organization level have been calculated using a modularity approach proposed by Newman (2006). This calculation procedure is able to find natural partitioning in dense networks. A block model approach to assess the structural equivalence would not be feasible, given the number of nodes at the organization level. The provinces in each partition are in descending order according to the number of organizations from the province. Only the most relevant provinces in each partition are displayed.